\documentclass[sigconf,anon,nonacm]{acmart}

\AtBeginDocument{%
  }

\setcopyright{acmcopyright}
\copyrightyear{2018}
\acmYear{2018}
\acmDOI{XXXXXXX.XXXXXXX}

\acmConference[Conference acronym 'XX]{Make sure to enter the correct
  conference title from your rights confirmation emai}{June 03--05,
  2018}{Woodstock, NY}
\acmPrice{15.00}
\acmISBN{978-1-4503-XXXX-X/18/06}

\usepackage{numprint}
\npthousandsep{,}

\usepackage{balance}
\usepackage{amsthm}
\usepackage[ruled, vlined, linesnumbered]{algorithm2e}
\usepackage{caption}
\usepackage{subcaption}
\usepackage{multirow}
\usepackage{multicol}

\usepackage{enumitem}
\setlist{leftmargin=3.5mm}

\usepackage{microtype}

\newcommand{\gitseed}[0]{\textsc{GitSEED}\xspace}
\newcommand{\GitLab}[0]{\textsc{GitLab}\xspace}
\newcommand{\cfaults}[0]{\textsc{CFaults}\xspace}

\begin{document}

\title{\gitseed: A Git-backed Automated Assessment Tool for Software Engineering and Programming Education}





\author{Pedro Orvalho}
\orcid{0000-0002-7407-5967}
\email{pmorvalho@tecnico.ulisboa.pt}
\affiliation{%
  \institution{INESC-ID, IST, U. Lisboa}
  \city{Lisboa}
  \country{Portugal}
}

\author{Mikoláš Janota}
\orcid{0000-0003-3487-784X}
\email{mikolas.janota@cvut.cz}
\affiliation{%
  \institution{Czech Technical University in Prague}
  \city{Prague}
  \country{Czech Republic}
}

\author{Vasco Manquinho}
\orcid{0000-0002-4205-2189}
\email{vasco.manquinho@tecnico.ulisboa.pt}
\affiliation{%
  \institution{INESC-ID, IST, U. Lisboa}
  \city{Lisboa}
  \country{Portugal}
}



\begin{abstract}
  
  Due to the substantial number of enrollments in programming courses, a key challenge is delivering personalized feedback to students. The nature of this feedback varies significantly, contingent on the subject and the chosen evaluation method. However, tailoring current Automated Assessment Tools (AATs) to integrate other program analysis tools is not straightforward. Moreover, AATs usually support only specific programming languages, providing feedback exclusively through dedicated websites based on test suites.

  This paper introduces \gitseed, a language-agnostic automated assessment tool designed for Programming Education and Software Engineering (SE) and backed by \GitLab. The students interact with \gitseed through \GitLab. Using \gitseed, students in Computer Science (CS) and SE can master the fundamentals of git while receiving personalized feedback on their programming assignments and projects. Furthermore, faculty members can easily tailor \gitseed's pipeline by integrating various code evaluation tools (e.g., memory leak detection, fault localization, program repair, etc.) to offer personalized feedback that aligns with the needs of each CS/SE course. 
  Our experiments assess \gitseed's efficacy via comprehensive user evaluation, examining the impact of feedback mechanisms and features on student learning outcomes. Findings reveal positive correlations between \gitseed usage and student engagement.
  
  
\end{abstract}





\maketitle

\section{Introduction}
\label{sec:intro}

The increasing need for programming and Software Engineering (SE) education has led to the emergence of various online courses, including Massive Open Online Courses (MOOCs)~\cite{clara,C-Pack-IPAs,InvAASTCluster-corr22}.
Providing feedback to CS students on their programming assignments and projects demands considerable time and effort from the faculty. Thus, there is a rising demand for systems, such as Automated Assessment Tools (AATs), that can deliver automated, comprehensive, and personalized feedback to students. 
When compared to non-automated evaluators, such as teaching assistants, AATs can evaluate several assignments or code submissions efficiently and quickly. Hence, AATs facilitate the learning process since students get their feedback much faster. Moreover, AATs offer objectivity and consistency, adhering to some evaluation metric (e.g., a test~suite). 

The interest and development of AATs dates back to the 1960s~\cite{acm60-1st-grader, tce22-AATs-survey}. 
Over the past two decades, there has been a surge in the growth and adoption of AATs~\cite{mooshak03,webcat08,autolab,iticse16-enki,sigcse17-submitty,iticse20-check50,codeboard.io}.
However, despite the remarkable growth in the development and usage of AATs, certain drawbacks have become increasingly apparent.
Primarily, a majority of AATs~\cite{mooshak03, codeboard.io} merely display the outcomes of a set of input/output tests used for the student's evaluation and lack other kinds of feedback. Secondly, ATTs tend to be specific to one programming language or a limited set of languages. AATs that are language-agnostic are scarce~\cite{webcat08,sigcse17-submitty}. Thirdly, AATs typically offer feedback solely through dedicated websites, necessitating students to familiarize themselves with new GUI interfaces.
Finally, it is either challenging or impractical to adapt most AATs to integrate other program analysis tools, which might be essential to provide more personalized feedback in some CS/SE courses.

\begin{figure*}[t!]
  \includegraphics[width=0.8\textwidth]{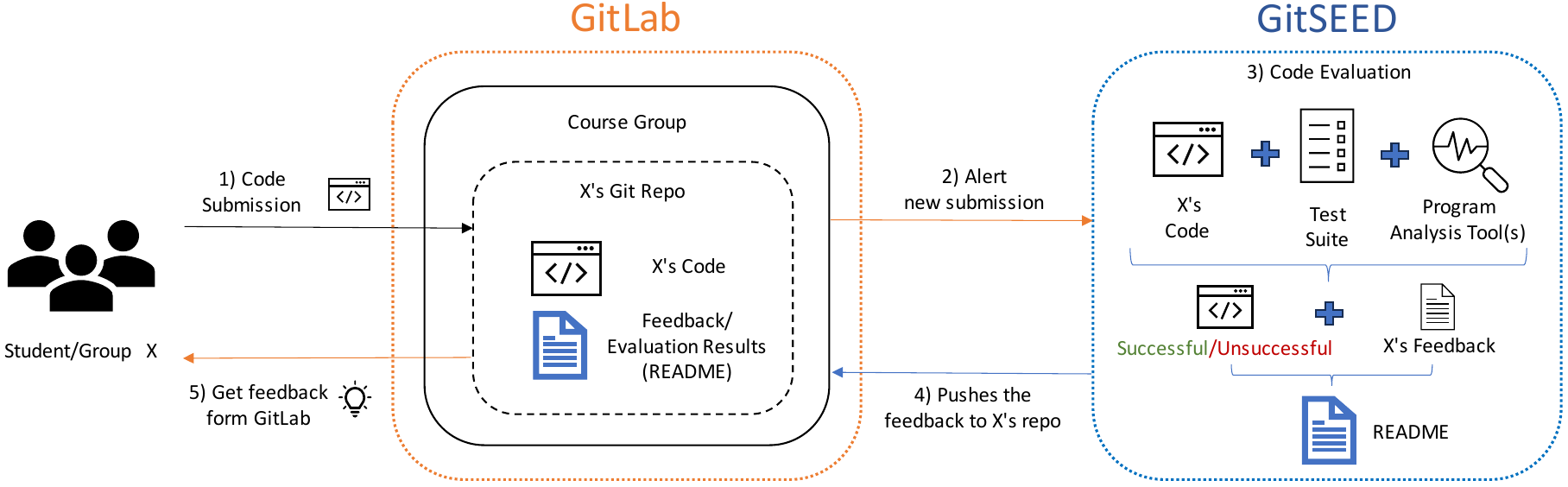}
  \caption{The overview of \gitseed.}
  \Description{Overview of \gitseed.}
  \label{fig:gitseed-overview}
\end{figure*}

This paper introduces \gitseed, a new tool that overcomes the aforementioned limitations of previous AATs. \gitseed is a novel \textbf{Git}-backed AAT for \textbf{S}oftware \textbf{E}ngineering and Programming \textbf{Ed}ucation. Figure~\ref{fig:gitseed-overview} presents the overview of \gitseed. 
As Figure~\ref{fig:gitseed-overview} shows, the students interact with \gitseed through \GitLab.
This way, CS/SE students can learn the fundamentals of git while receiving personalized feedback on their programming assignments and projects. Afterwards, using \GitLab's runners, \gitseed is notified whenever there is a new submission from group \textsc{X}. \gitseed evaluates this new submission against a test suite and using program analysis tools defined by the faculty. Finally, the resulting evaluation report is pushed into \textsc{X}'s git repository (repo) so that the students have access to personalized feedback right away.

Furthermore, \gitseed is language agnostic, i.e., it can be used for any CS/SE course no matter the programming language(s) used.
Moreover, most CS/SE students are familiar or will be familiar throughout their courses, with several git web interfaces (e.g., \GitLab, \textsc{GitHub}, \textsc{Gitea}). Therefore, \gitseed eliminates the necessity for students to acquaint themselves with an unfamiliar GUI interface, which happens frequently in several universities where different CS/SE courses use different GUI interfaces for automated assessment of programming tasks~\cite{tce22-AATs-survey,autolab,codeboard.io,webcat08,mooshak03}. 

\gitseed has two different categories of assessments: labs and projects. Either one is optional, and it is possible to have an unlimited number of projects depending on the chosen configuration. Faculty can choose which assessment model aligns best with their courses. 
Moreover, faculty members can easily tailor the pipeline of \gitseed, enhancing the quality of feedback provided to the students aligned with the needs of each course. For example, code evaluation tools can be integrated into \gitseed, such as memory leak detection~\cite{nethercote2003valgrind}, fault localization~\cite{Cfaults-arXiv24}, 
program repair~\cite{clara,ecai23-GNNs-4-var-mapping}, plagiarism detection~\cite{moss}, code coverage~\cite{gcov}, among~others~\cite{master-thesis-pedro,cp19-enum-PS}. 

The paper is organized as follows. Section~\ref{sec:gitseed} presents the implementation and possible configurations of \gitseed in more detail. 
\gitseed has already proven successful in two separate courses, a first-year programming course and a CS graduate course. Section~\ref{sec:impact} discusses our experiments and evaluates the effectiveness of \gitseed in enhancing programming education. Through the analysis of feedback from students enrolled in a first-year undergraduate course, we explore the role of \gitseed's features, including dashboards and feedback mechanisms, in facilitating learning and improving student performance.
Finally, Section~\ref{sec:related} briefly reviews related work, and the paper concludes in Section~\ref{sec:conclusion}. 

To summarize, this paper makes the following contributions:
\begin{itemize}
    \item We present \gitseed, an open-source language-agnostic automated assessment tool designed for Software Engineering (SE) and Programming Education and backed by \GitLab;
    \item \gitseed is integrated into \GitLab's continuous integration (CI) workflow, which adopts educational assessment within a professional version control platform rather than a dedicated website, like so many other AATs;
    \item Students interact with \gitseed through \GitLab, learning this way the fundamentals of git while receiving personalized feedback on their assignments;
    \item Faculty members can easily adapt \gitseed by integrating other code analysis tools to offer personalized feedback that aligns with the needs of each CS/SE course.
    \item \gitseed is publicly available on GitLab~\cite{GitSEED-GitLab-2024} and on \textsc{Zenodo}~\cite{GitSEED-Zenodo-2024}.
\end{itemize}


\section{\gitseed}
\label{sec:gitseed}

This section presents the internals of \gitseed and configuration options.
Section~\ref{sec:gitlab} describes the \GitLab features required by \gitseed.
Next, Section~\ref{sec:vm} focuses on the back-end of \gitseed and its workflow, and Section~\ref{sec:security} details the measures taken in order to ensure that all stages of the \gitseed pipeline are safe. Finally, Section~\ref{sec:implementation} explains our implementation of \gitseed.

\begin{figure*}[t!]
  \includegraphics[width=0.75\textwidth]{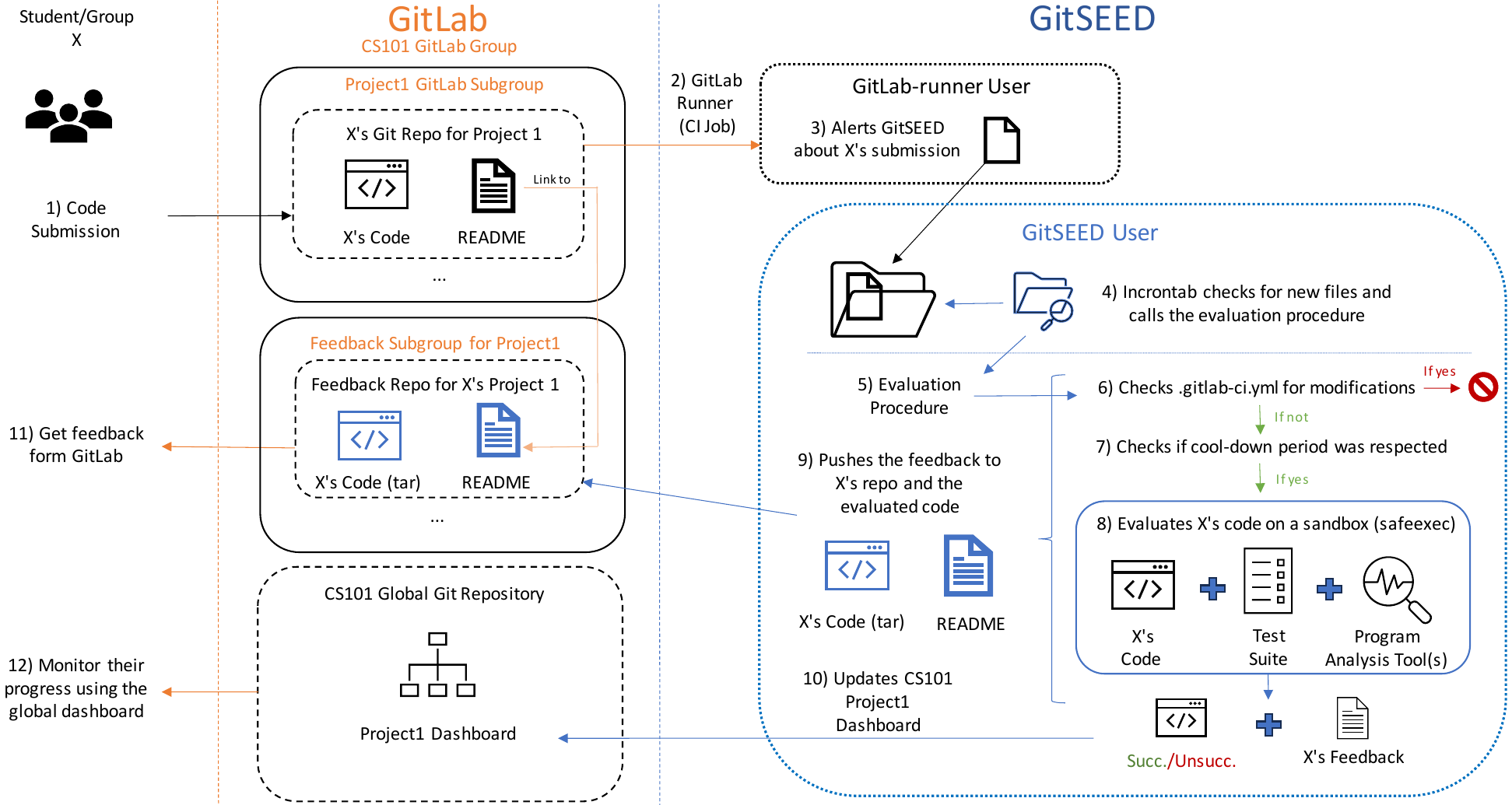}
  \caption{Workflow of \gitseed for processing a new project submission from group X in CS101.}
  \Description{The workflow of \gitseed.}
  \label{fig:gitseed-workflow}
\end{figure*}

\subsection{\GitLab}
\label{sec:gitlab}

Figure~\ref{fig:gitseed-workflow} illustrates the complete workflow of \gitseed, where \GitLab plays the role of an intermediary between the students and \gitseed.
Notice that students only interact with \GitLab, and then \GitLab triggers the processing of submissions in \gitseed.
Furthermore, \gitseed was designed to work with other git web interfaces. 
Modern, widely-used programming editors/environments (e.g., \textsc{VS Code}) already feature user-friendly interfaces for managing git repos. Hence, the submission process and getting feedback can be easily done within the student's programming environment, eliminating the need to exit their coding workspace. Alternatively, students can also use \GitLab's web interface. 

\subsubsection{GitLab Group, Subgroups and Course Repo}

\gitseed expects the following structure of groups in \GitLab: a main \GitLab group with all the students (e.g., CS101), a subgroup for each distinct evaluation element (e.g., labs, project, etc.), an additional subgroup for feedback, and a git repo for the entire course (CS101, in Figure~\ref{fig:gitseed-workflow}) that contains all the course's information and dashboards for each evaluation element.

\gitseed has different repos for labs and projects. The rationale behind this choice is that while labs remain open throughout the semester, projects have distinct deadlines and may involve different groups of students.
Furthermore, the feedback can also be pushed directly to the same git repos used by students for code submissions. However, having two different repos, one for code submissions and one for getting feedback, is the best approach for first-year students. This approach mitigates the chances of merging conflicts or git conflicts between students and \gitseed. Hence, \gitseed uses different repos to simplify the students' repos synchronization. This way, students manage their code development repos, and \gitseed only submits to the feedback repo.

In \GitLab, each project member is assigned a role that determines which actions they can take in the git repo~\footnote{\url{https://docs.gitlab.com/ee/user/permissions.html}}.
In \gitseed, students assume the role of ``Developers'' for their own repos (e.g., projects, labs), granting them read and write access. However, students assume the role of ``Reporters'' for their feedback git repos and in the course global project, granting them viewing but not editing privileges. Note that the group of students can only see their own feedback repo and not those of other groups.
Finally, faculty members hold the roles of ``Maintainers'' or ``Owners'' for all repos, depending on the chosen configuration.

\subsubsection{Continuous Integration (CI)}%
\label{sec:CI}

\gitseed takes advantage of the CI pipeline\footnote{\url{https://docs.gitlab.com/ee/ci/}} available on \GitLab. 
The CI pipeline, essential for testing and deploying software projects, operates through a \texttt{.gitlab-ci.yml} script outlining all testing and deployment actions. It utilizes ``runners'', agents executing these actions, such as tests, as defined in the script.

\paragraph{GitLab Runner} \gitseed requires a \GitLab runner to be installed and self-hosted in the machine where \gitseed is running (e.g., a Linux virtual machine (VM)).
In this machine, the runner runs the job described in the \texttt{.gitlab-ci.yml}. The runner's job is to add information to \gitseed's queue that student X made a new submission to project Y. 
The faculty needs to adapt the \texttt{.gitlab-ci.yml} based on the programming language(s) being evaluated in the course. Distinct students can use different programming languages for the same programming task.

\subsection{\gitseed's Back-end}%
\label{sec:vm}

Next, we detail the stages of the \gitseed workflow from Figure~\ref{fig:gitseed-workflow}.

\subsubsection{\GitLab Manager} The \GitLab manager interacts with \GitLab and creates/modifies/clones all the necessary subgroups and repos and assigns the students and faculty to their own repos. \gitseed works for single-student groups and for groups with several students. Using the \GitLab manager, the faculty can easily manage students' access to their git repos. The students' accesses can be removed, for example, after a project deadline or if the students modified something in the git repo that were not supposed to (e.g., \texttt{.gitlab-ci.yml}).

\subsubsection{Assessments} \gitseed has two categories of assessments: labs and projects. Either one is optional, and it is possible to have an unlimited number of projects depending on the chosen configuration. Each type of assessment has its own evaluation script.

\paragraph{Labs} \gitseed allows faculty members to publish each different lab's exercises in the student's repos. Each lab corresponds to a practical class. Additionally, with \gitseed, students who do not finish all the lab exercises during the class can still conclude and automatically check their implementations afterwards.

\paragraph{Projects} \gitseed allows the publication of the projects' descriptions and related data in the students' repos. Moreover, after a project's deadline, faculty members can remove the students' access to write into their repos and then reevaluate all the projects one last time. This last reevaluation might be necessary in case there was any submission that was not assessed due to the \emph{cool-down period} (see description of cool-down periods next).

\subsubsection{Commits Database} \gitseed maintains a database containing all students' commits timestamps for every evaluation element. The goal is to have a cool-down period for each different evaluation element, ensuring that only the submissions respecting their previous cool-down period are evaluated. This measure is implemented to prevent overloading \gitseed's machine. 
Note that some CS/SE courses have thousands of students, who tend to submit multiple times, especially near project deadlines. Moreover, a cool-down period forces students to think more thoroughly about their program before making new submissions, as no new feedback will be provided for commits made during this time. The default cool-down period is set at 1 minute for lab exercises and 10 minutes for project assignments. Nevertheless, these periods can be easily modified (see Section~\ref{sec:configs}). Furthermore, the feedback report (README) lets the students know when their current cool-down period is over.

\subsubsection{Dashboard} \gitseed keeps a dashboard/leaderboard for each distinct evaluation element. These dashboards are automatically posted by \gitseed in the course's central git repo so all students can monitor their progress with regard to their colleagues. The dashboards keep track of each student/group's number of successful/unsuccessful tests, their number of submissions, and the number of days since the beginning of the project/lab assignment.

\subsubsection{\GitLab-Runner} The runner (see Section~\ref{sec:CI}) needs to be installed in the same machine where \gitseed is running and to have write access to the folder that keeps track of new submissions.

\subsubsection{Incrontab} The machine where \gitseed is installed has an incrontab daemon that is triggered by the \GitLab runner. The \GitLab runner adds the information that a new commit was performed on a given repo, and that triggers the evaluation procedure.

\subsubsection{Evaluation} During the evaluation process, \gitseed first checks if the students modified the \texttt{.gitlab-ci.yml} script. If this is the case, students lose their rights to push/modify the git repo and are asked to reach out to faculty members. 
Otherwise, \gitseed checks if the cool-down period was respected. If not, the student's new submission is not evaluated. Otherwise, if the cool-down period was respected, then \gitseed proceeds to the next evaluation step.

\paragraph{\texttt{safeexec}} To run the students' code safely, \gitseed uses \texttt{safe\-exec}~\cite{safeexec} which is a lightweight sandbox for executing user programs. Alongside \texttt{safeexec}, other program analysis tools can be run on the students' code.
After completing the evaluation, \gitseed submits the evaluation report (README) and a tar file containing the evaluated code (e.g., a programming assignment's implementation) to the respective feedback repo. Finally, \gitseed updates the respective dashboard in the course's central git repo with the student's performance. 

\subsubsection{Configurations}%
\label{sec:configs}
\gitseed has several predefined configurations that can be easily modified in the configuration file:

\begin{itemize}
    \item Cool-down Period (default: 1 min): Amount of time students need to wait between their own submissions;
    \item Output Visible (default: false): \gitseed shows (or does not) the output of all tests to the students;
    \item Only First Wrong Output Visible (default: true): \gitseed only shows students their first incorrect output. This option is only used if the previous option is set to true;
    \item CPU Time Limit (default: 5 sec): \gitseed runs the students' code with this CPU time limit for each test case;
    \item Memory Limit (default: 8 GB): \gitseed runs the students' code with this memory limit for each test case.
\end{itemize}

\paragraph{Easily Tailored}
\gitseed's current methods for evaluation are fully language agnostic. The evaluation scripts can be easily tailored to evaluate different programming languages.
Furthermore, faculty members can quickly adapt the \gitseed pipeline by integrating or replacing various code evaluation tools (e.g., memory leak detection, fault localization, program repair, plagiarism checks, solution checkers) to offer personalized feedback that aligns with the needs of each CS/SE course. 
Lastly, \gitseed was designed with modularity in mind. On that account, one can easily remove, add, or modify any component of \gitseed without compromising it.

\subsection{Safety Measures}%
\label{sec:security}

Several measures must be ensured for \gitseed to operate safely.
Firstly, the \GitLab runner user needs to be granted write access to \gitseed's folder for new submissions. However, this user should not have access to any other folders, as the \GitLab runner executes code from the \texttt{.gitlab-ci.yml} script, which may be tampered with by students. By limiting access to only that folder, the \GitLab runner cannot alter or read anything else from \gitseed.
Furthermore, \gitseed runs the students' code using \texttt{safeexec}, which simulates a sandbox controlling read/write accesses. Note that \gitseed is not dependent on \texttt{safeexec}. Due to \gitseed modularity, \texttt{safeexec} can be quickly replaced with some other sandbox application.
Lastly, given the crucial role of \texttt{.gitlab-ci.yml} in \gitseed's functionality, this script is added to the \texttt{gitignore} file and students' READMEs explicitly instruct them not to edit this \texttt{yml} script. Nevertheless, \gitseed checks for any tampering with this script before evaluating the student's code. If detected, \gitseed restricts the student's access to the repo until the faculty checks the situation.

\subsection{Implementation}%
\label{sec:implementation}

For this paper, we used a \GitLab instance self-hosted at 
Instituto Superior Técnico.
The \gitseed system was deployed on a dedicated virtual machine running Linux (Debian 4.19) on a AMD Opteron(TM) Processor 6276 with 16GB of RAM. Additionally, the virtual machine hosted the gitlab-runner package, version 15.9.1.
\gitseed  is implemented using \texttt{bash} and python3 (version 3.9.16). \gitseed uses \texttt{bash} to execute and evaluate the students' code. On the other hand, \gitseed relies on python3 and \texttt{curl} to communicate with \GitLab, through its API (v3.15.0). Furthermore, \gitseed utilizes python3 and \texttt{sqlite3} for the maintenance of the course's dashboards and the database containing the commit history.

\section{Impact Discussion}%
\label{sec:impact}

This section discusses our experiments with \gitseed, between Spring 2023 and Spring 2024, in two distinct academic courses at 
Instituto Superior Técnico, 
a first-year undergraduate and a graduate course.
\gitseed offers both formative and summative assessments. Formative assignments, such as lab classes, remain accessible throughout the semester, allowing students to revise until correct. Summative assignments, such as projects, also permit unlimited attempts but come with strict deadlines.
Students were briefed that formative assignments serve as learning aids, encouraging exploration without fear of repercussions for errors. The aim is for students to utilize \gitseed until mastery is achieved. Conversely, summative assignments serve as assessments of acquired knowledge and skill, showcasing proficiency in the subject. Since these summative assignments require more computation time and memory, higher cool-down periods were established between each group's submissions, to prevent overloading \gitseed's machine.

\subsection{Courses Setup}

\subsubsection{\textbf{Undergraduate Course (Spring 2023)}}%
\label{sec:iaed-1st}
 \gitseed was initially used in Spring 2023 in a first-year undergraduate course, Introduction to programming in C, with a total of 528 enrolled students. \gitseed was used to assess this course's labs (formative assignments) and projects (summative assignments). 

 \paragraph{Assignments} For formative assignments, \gitseed created individual git repos for the lab classes. Configuration settings included a one-minute cool-down period, a five-second CPU time limit, and an 8GB memory limit for each programming assignment across eight labs. Throughout these eight labs, students made a total of 10338 code contributions to their repos. 
 Regarding the summative assignments, this course had two different programming projects, each configured for single-student groups. Configuration settings included a five-second CPU time limit, a 16GB memory limit, and a 10-minute cool-down period for project evaluations. Students made a total of 15061 code contributions to their repos, 7916 to the first project and 7145 to the second one. 
 
 \paragraph{Program Analysis Tools} For projects evaluation, \gitseed was tailored to: (1) identify forbidden libraries in student projects, and (2) run \textsc{valgrind}~\cite{nethercote2003valgrind} to detect memory leaks in their code. Providing feedback on memory leaks proved beneficial, particularly for first-year students unfamiliar with these tools.
 
\paragraph{Opportunities for improvement} Throughout the course, we noticed that some students forgot to pull the feedback from \GitLab to their local repos before pushing new modifications, resulting in git-merge issues. Consequently, \GitLab's CI would ignore these commits. This happened because students use \GitLab web interface to get their feedback while coding in their local git repos. To address this issue, \gitseed now publishes feedback in separate repositories, with students' READMEs containing links for easy synchronization, especially for first-year students.

\subsubsection{\textbf{Graduate Course (Fall 2023)}}
 In Fall 2023, \gitseed was also used in a graduate course on Automated Reasoning with 38 students. 
 The project consisted of solving an NP-Hard optimization problem through a Boolean logic solver using Python.
 There were 21 groups, each consisting of a maximum of two students. Configuration settings included a one-minute CPU time limit, a 16GB memory limit, and a 20-minute cool-down period for project evaluations. Throughout the project, students made a total of 269 contributions to their repos. Once again, we customized \gitseed, in this case, to incorporate both private and public test cases for evaluating students' code. Moreover, we also inserted additional software into the \gitseed pipeline that gave students feedback about the satisfiability and optimality of their projects' solutions.

\subsubsection{\textbf{Undergraduate Course (Spring 2024)}}

In Spring 2024, \gitseed once again played a pivotal role by supporting the first-year undergraduate course, Introduction to Programming in C, which had a total enrollment of 509 students. \gitseed served as the platform for assessing labs and projects in the course, employing configurations similar to those outlined in Section~\ref{sec:iaed-1st}. However, notable adjustments were made for this course iteration.

\paragraph{Feedback}
Feedback was provided in separate repositories based on the insights gained from the previous year.

\paragraph{Program Analysis Tools}
A significant enhancement was the integration of four additional program analysis tools into \gitseed: \cfaults,
\textsc{cppcheck}, \textsc{clang-tidy}, and \texttt{Lizard}. \texttt{Lizard}~\cite{lizard} is a cyclomatic complexity analyzer for various programming languages, aiding in evaluating code length and complexity. The fault localization tool pinpointed faulty statements within the programs based on a test suite. Additionally, \textsc{cppcheck}~\cite{cppcheck} and \textsc{clang-tidy}~\cite{clang-tidy} are static analyzers used to detect uninitialized variables and various errors, such as division by zero. Finally, \cfaults~\cite{CFaults-FM24} is a formula-based fault localization tool that pinpoints bug locations within the programs. The insights generated by these tools were compiled into feedback reports and appended alongside test-suite evaluation outcomes in the students' feedback repositories. The results from both the fault localization tool and the static analyzers were presented to students as ``Hints'', strategically guiding them towards potential problematic statements within their programs. This approach aimed to provide students with targeted assistance in identifying and rectifying programming errors.
Moreover, \gitseed was configured to display only the first incorrect output to students, fostering a focused learning environment.

\subsection{User Study}

In Spring 2024, we conducted a comprehensive user study with students to gather valuable feedback on their experience with \gitseed, particularly focusing on its dashboards and the various types of feedback provided by analysis tools, namely \texttt{valgrind}, \texttt{lizard}, and ``hints'' (generated by fault localization and static analyzers).
Throughout the course, we noticed that incorporating motivational elements, such as the dashboards available within \gitseed, effectively encouraged student engagement and facilitated their progress.
Approximately 20\% of the students who were enrolled in the course for the entire semester took part in the questionnaire. They were asked anonymously to evaluate the usefulness of the different feedback mechanisms and features of \gitseed they encountered during the semester (see Appendix~\ref{appendix}).
The findings revealed that students perceived the following aspects as particularly beneficial:

\textbf{\gitseed}: 91.8\% of students found \gitseed to be a valuable resource. Its role in providing a centralized platform for assignment submission, feedback reception, and revision evidently streamlined the learning process and enhanced overall comprehension.
\textbf{Dashboards}: 82.2\% of students acknowledged the significance of dashboards in tracking their progress and monitoring their performance relative to course objectives. 
\textbf{Hints}: Despite being less prevalent than other feedback mechanisms, 68.5\% of students recognized the utility of hints generated by fault localization and static analyzers. These hints acted as invaluable pointers, directing students towards potential errors in their code and fostering a deeper understanding of programming concepts through self-correction.
\textbf{\texttt{Valgrind}}: 90.4\% of students found the feedback from \texttt{valgrind} to be beneficial. This tool's ability to detect memory management issues and provide detailed diagnostics undoubtedly aided students in debugging their programs and writing more robust code.
\textbf{\texttt{Lizard}}: 75.3\% of students appreciated the insights offered by \texttt{lizard}, particularly its analysis of code complexity and length. By highlighting areas of code that might require simplification or restructuring, \texttt{lizard} contributed to the optimization of students' coding practices and the cultivation of clearer, more efficient programming habits.
In addition to evaluating the specific components of \gitseed, students were given the opportunity to provide general feedback through short-answer responses. These open-ended questions allowed students to express their thoughts, suggestions, and concerns regarding their overall experience with \gitseed.
Overall, the user study underscored the positive impact of \gitseed's features and feedback mechanisms on students' learning experiences, reaffirming its value as a comprehensive educational tool for programming courses.

\section{Related Work}%
\label{sec:related}

\paragraph{Automated Assessment Tools (AATs)} Over the past decades, there has been a growing interest in the automated evaluation of Software Engineering (SE) and Computer Science (CS) students~\cite{tce22-AATs-survey}. Typically, AATs assess programming tasks using input/output (IO) tests predefined by the course's faculty. There exists a substantial number of AATs that function as web-based Integrated Development Environments (IDEs) for evaluating students' code using IO tests. Examples include \textsc{CodeOcean}~\cite{geec16-codeOcean}, \textsc{Mooshak}~\cite{mooshak03}, and \textsc{Web-Cat}~\cite{webcat08}.
\textsc{Enki}~\cite{iticse16-enki} is also a web-based IDE although it offers other kinds of pedagogical skills to the students, e.g., gamification features.
Furthermore, \textsc{Autolab}~\cite{autolab} and \textsc{Submitty}~\cite{sigcse17-submitty} are open-source web-based course management platforms that automatically grades students' code. \textsc{Autolab} maintains scoreboards for each evaluation element in order to motivate the students, while \textsc{Submitty} provides an interface for Teaching Assistants (TAs) to manually grade assignments.
Additionally, Codeboard.io~\cite{codeboard.io} is a web-based IDE to teach programming tasks. Faculty members can share programming exercises with the students. These exercises are assessed using a result string or a set of predefined unit tests. The set of programming languages available is limited and Codeboard.io is difficult to tailor in order to get more personalized feedback for the students.
\textsc{GradeStyle}~\cite{ics23-seet-GradeStyle} serves as a code style marker tool that provides feedback for assignments in Java by opening a GitHub issue in each student’s repository.
Moreover, \textsc{gradescope}~\cite{gradescope} is another online tool to administer and grade programming assignments as well as other kinds of assessments (e.g., exams). However, \textsc{gradescope} only has paid licenses for education.

\textsc{GitHub Classroom}~\cite{GitHubClasroom} is an AAT tool available on \textsc{GitHub} that allows faculty to create and manage digital classrooms and assignments. \textsc{GitHub Classroom} uses the same mechanism of a CI runner (GitHub Actions) to process student code and report on quality aspects. \textsc{GitHub Classroom}  shares several benefits with \gitseed, is language agnostic, and enables students to learn the fundamentals of git while receiving feedback on their assignments. However, \textsc{GitHub Classroom} operates on a third-party platform. There may be regulatory or institutional policies that restrict the use of cloud-based services for certain types of data. On the other hand, \GitLab is open-source and can be self-hosted by educational institutes.
Additionally, \textsc{GitHub Classroom} may not seamlessly integrate with existing learning management systems (LMS) used by educational institutions. This lack of integration can result in administrative challenges, such as maintaining separate platforms for course materials, grades, and communication. While \gitseed can be easily integrated with this kind of systems.
Finally, while \textsc{GitHub Classroom} is free to use, some advanced features or integrations may require paid \textsc{GitHub} plans. For example, the number of minutes available for GitHub Actions (CI) is limited per month. Instructors may need to consider the cost of providing \textsc{GitHub} accounts or repositories for students, especially in cases where institutional resources are limited. In contrast, educational institutions can use the premium version of \GitLab for free, and both \GitLab and \gitseed are open-source projects. Finally, it is worth noting that there is no equivalent to \textsc{GitHub Classroom} on \GitLab, highlighting an opportunity for \gitseed to fill this gap.

\paragraph{Competitive Programming Contests (CPCs)} CPCs are online platforms that host programming contests. In these websites, students and CS/SE professionals, engage in solving computational problems under time/memory constraints. These contests serve as platforms to assess and enhance problem-solving skills, algorithmic efficiency, and programming proficiency. Typically, CPCs assess contestants’ code using an IO test suite. \textsc{LeetCode}~\cite{leetcode},
\textsc{topcoder}~\cite{topcoder},
\textsc{Codeforces}~\cite{codeforces}, and
\textsc{replit}~\cite{replit} are among the most famous CPCs.
Both \textsc{Codeforces}~\cite{codeforces} and \textsc{replit}~\cite{replit} offer features for programming education. However, the evaluation process relies solely on IO tests.

\section{Conclusion}%
\label{sec:conclusion}

This paper presents \gitseed, an open-source, language-agnostic automated assessment tool (AAT) seamlessly integrated with \GitLab. Students benefit from personalized feedback on programming assignments and projects, mastering Git fundamentals simultaneously. Notably, \gitseed eliminates the need for students to navigate new GUI interfaces. Integrated into \GitLab's continuous integration (CI) workflow, \gitseed brings educational assessment into a professional version control platform rather than a dedicated web-based platform. Furthermore, faculty can easily customize \gitseed's pipeline with various code evaluation tools. Our experiments showcased \gitseed's success in both undergraduate and graduate courses, affirming its efficacy in programming education. It enhances student engagement and learning outcomes. Positive student feedback highlights \gitseed contribution to active learning and a supportive educational environment.



\begin{acks}
This work was partially supported by Portuguese national funds through FCT, under projects UIDB/50021/2020 (DOI: 10.54499/\-UIDB/\-50021/\-2020), PTDC/\-CCI-COM/\-2156/2021 (DOI: 10.54499/\-PTDC/\-CCI-COM/\-2156/\-2021) and 2022.\-03537.PTDC (DOI: 10.54499/\-2022.03537.PTDC) and grant SFRH/\-BD/\-07724/\-2020 (DOI: 10.54499/\-2020.07724.BD). 
PO acknowledges travel support from the EU’s Horizon 2020 research and innovation programme under ELISE Grant Agreement No 951847.
This work was also supported by the MEYS within the program ERC CZ under the project POSTMAN no.~LL1902 and co-funded by the EU under the project \emph{ROBOPROX} (reg.~no.~CZ\-.02.01.01/00/\-22\_008/0004590). This article is part of the RICAIP project funded by the EU’s Horizon 2020 research and innovation program under grant agreement No 857306.
\end{acks}


\bibliographystyle{ACM-Reference-Format}
\balance
\bibliography{mybibliography}

\appendix

\section{User Study Questions}
\label{appendix}

\begin{itemize}
    \item \textbf{Q1.} In your opinion, was the use of the code submission/evaluation system in this course helpful?
    \item \textbf{Q2.} In your opinion, were the dashboards for the labs and the project helpful?
    \item \textbf{Q3.} In your opinion, were the hints provided in the feedback for Lab 2 submissions helpful?
    \item \textbf{Q4.} In your opinion, was the Valgrind report in the feedback for the labs and project helpful?
    \item \textbf{Q5.} In your opinion, was the Lizard report in the project feedback helpful?
    \item \textbf{Q6.} If you would like to leave any comments or suggestions, please enter them here.
\end{itemize}


\end{document}